# Stern-Gerlach detection of neutral atom qubits in a state dependent optical lattice

Tsung-Yao Wu, Aishwarya Kumar, Felipe Giraldo Mejia, David S. Weiss[*]

Department of Physics, The Pennsylvania State University, University Park, PA 16802

Qubit state measurements are an essential part of any quantum computer, constituting the readout. Accurate measurements are also an integral component of one-way quantum computation and of error correction, which is needed for fault tolerant quantum computation[1]. Here we present a state measurement for neutral atom qubits based on coherent spatial splitting of the atoms' wavefunctions. It is reminiscent of the Stern-Gerlach experiment[2], but carried out in light traps. For ~160 qubits in a 3D array, we achieve a measurement fidelity of 0.9994, which is ~20 times lower error than in any previous measurement of a neutral atom array[3,4]. To our knowledge, it also significantly exceeds the demonstrated measurement fidelity of other arrays with more than four qubits, including those with ion and superconducting qubits[5,6] Our measurement fidelity is essentially independent of the number of qubits measured, and since the measurement causes no loss, we can reuse the atoms. When atoms have been lost to background gas collisions during the experiment, we demonstrate here that we can replace them[7].

Neutral atoms are promising qubit candidates, given that they are identical, can be readily scaled to large arrays, and have long coherence times[8]. A simple way to measure the states of atom qubits is to resonantly push (clear) atoms in one qubit state out of the trap with light and detect those that remain. The detection fidelity for the remaining atoms can exceed 0.9997[9], but the scheme suffers from two considerable drawbacks. First, qubit loss during the computation is indistinguishable from the atom being in the cleared state, making the loss rate the de facto limit on the state measurement fidelity. Second, about half the atoms are cleared during measurement, necessitating reloading of atoms, thus making it hard to adapt this method to error correction. A few alternative approaches have been taken to achieve lossless state detection. Atoms in each state in turn can be made to fluoresce on a cycling transition[3,4,10,11]. However, since the atoms cannot be cooled during the measurement, this method is a balance between detecting enough photons to identify the atom's state and those photons heating the atom out of the trap. The best results on small 1D atom ensembles have had 0.987 fidelity and 2% heating loss[3] or 0.98 fidelity and 1% heating loss[4]. Photon collection efficiency improves when a single atom is trapped in a high finesse cavity[12,13], where 0.9992 fidelity has been achieved[13], but this enhancement is hard to scale to more atoms. In a quantum gas microscope, a large magnetic field gradient was applied to coherently separate atoms in two different internal states, reminiscent of the Stern-Gerlach experiment, after which the two states were trapped and detected at adjacent lattice sites[14]. A fidelity of 0.98 was reached, limited by lattice phase fluctuations. Our technique is also a Stern-Gerlach type approach, but without gradient magnetic fields.

[*] Corresponding Author. dsweiss@phys.psu.edu

Our state detection scheme is conceptually illustrated in Fig. 1a. We start with atoms in a 3D lattice with a large spacing in all directions (lattice XYZ) in an unknown superposition of two internal states. To detect each atom's state, we adiabatically transform the state-independent X lattice into two state-dependent potentials that move in opposite directions[7,15,16]. The two state components of each atom follow their respective potentials, spatially splitting the wavefunction in two. Next, we replace $X$ with $X_S$, which has an order of magnitude shorter lattice spacing. The two parts of the wavefunction are each localized to within a couple of $X_S$ lattice sites. We then image the atoms with cooling light[9], which projects the wavefunction of each atom to a single site in $X_S YZ$, and measures its location. Mapping the internal state onto spatial position in this way avoids the detection/heating tradeoff that has limited previous measurements.

The $X$, $Y$, and $Z$ lattices in our experiment are created by pairs of 839 nm laser beams, crossing at 10° angles as shown in Fig. 1b, linearly polarized perpendicular to their plane of propagation. The lattice beams are slightly mutually shifted in frequency (by tens of MHz), and together form an approximately cubic 3D lattice with 4.8 μm spacing. The vibrational frequency of an atom trapped near the bottom of a lattice site is 15 kHz[9]. Cs atoms are loaded into the lattice from a magneto-optic trap. We can either use the approximately 40% random occupancy we start with, or we can sort atoms to fully fill a sub-lattice[7]. We detect atoms in five planes by imaging phase-scrambled (see Methods) polarization gradient cooling light (PGC) one $Z$ plane at a time, which takes 830 ms total (see Methods). Projection sideband cooling[7,17] leaves 89% of the atoms in their 3D vibrational ground states and >99.9% of them into the $|F = 4, m_F = -4\rangle$ hyperfine ground state, where $F$ and $m_F$ are the hyperfine and magnetic quantum numbers, respectively. We can transfer atoms into any other magnetic sublevel with a series of adiabatic fast passage (AFP) microwave pulses [18], and we can create superpositions of sublevels using Blackman pulses[18].

The 3D lattice initially traps all hyperfine sublevels nearly identically (see Fig. 1a-i). We make the $X$ lattice sublevel-dependent by rotating the polarization of one of its beams, which we accomplish with two Pockels cells and a $\lambda/4$ plate[7]. Atoms in different hyperfine levels with the same sign $m_F$, like $|F = 4, m_F = -4\rangle$ and $|F = 3, m_F = -3\rangle$, move in opposite directions. A nearly $\pi/2$ polarization rotation separates the two states by almost half of the $X$ lattice spacing (see Fig. 1a-ii). Ramping the voltage on the Pockels cell in 300 μs keeps the motion adiabatic. At the end of the ramp we suddenly turn on a retro-reflected ~150 μm waist laser beam that forms a lattice, $X_S$, with 0.42 μm spacing (see Figs. 1b and 1a-iii). To avoid mutual interference, the wavelength of the $X_S$ light is slightly different (by tens of MHz) from the other lattice beams. The spatial phase of $X_S$ relative to $X$ is not controlled. After the initial turn on, the $X_S$ power is increased in 78 μs so that the vibrational frequency in the ground state is raised from 43 kHz to 98 kHz, a sequence empirically adjusted to avoid site hopping in $X_S$. We then turn off $X$ and turn on the PGC light to image atoms. The atom's wavefunction is thus projected into a single site of the $X_SYZ$ lattice (see Fig. 1a-iv).

Figs. 2a shows the results of measurements starting with all atoms in $|F = 4, m_F = -4\rangle$ after optical pumping. For Fig. 2b, the atoms start in $|F = 3, m_F = -3\rangle$ after an AFP pulse and a $F=4$ state clearing laser pulse to ensure clean state preparation. The single plane images are typical and the grid vertices demark the initial atom locations. One can clearly see the detected atoms shifted to the left for $|F = 4, m_F = -4\rangle$ and to the right for $|F = 3, m_F = -3\rangle$. The locations of the atoms are fit to a floating $x$ center (see Methods), and the distributions of these positions are shown in the histograms, which quantify the qualitative shifts visible in the images. For Fig. 2c, the atoms start in a superposition of the two internal states. We set the line that separates atoms in the two states at -0.2 µm, equidistant from the two peaks, and color the histograms and the occupancy maps in the Figure accordingly.

The Gaussian root mean square widths of the displacement distributions in Fig. 2 are 203 nm, which is about half an $X_S$ lattice spacing (210 nm). It mostly results from the random phase relationship between $X$ and $X_S$. We infer that the asymmetry of the displacement centers results from slight imperfections in the $X$ polarization (see Methods). Across the entire 9 × 9 × 5 array, 7 $\binom{+4}{-3}$ × 10$^{-4}$ of the atoms nominally prepared in $|F = 4, m_F = -4\rangle$ are measured to be in $|F = 3, m_F = -3\rangle$ and 5 $\binom{+3}{-2}$ × 10$^{-4}$ of the atoms nominally prepared in $|F = 3, m_F = -3\rangle$ are measured to be in $|F = 4, m_F = -4\rangle$ (see Fig. 2a and 2b insets), yielding an average state detection fidelity of 0.9994.

When atoms spontaneously emit lattice light before they are captured in $X_S$, they can change states and move in the wrong direction. We calculate a scattering rate of 3 × 10$^{-4}$ during the motion in $X$, which is consistent with this being the dominant source of error. If the $X$ polarization were to be improved, we could move the atoms faster and thus reduce the scattering and the associated state detection error. We could also employ a "throw and catch" method. In this scheme, we would suddenly change an $X$ beam polarization, let the atoms be accelerated by the shifted lattices, and then shut off $X$ to avoid spontaneous emission. When the atoms have traveled sufficiently far, $X_S$ would be turned on with cooling light to catch the atoms. We estimate that this could reduce our error from spontaneous emission by a factor of five.

After the state measurement, we adiabatically turn X back on with its standard polarization and then turn off $X_S$ with the essentially reverse sequence we used to turn it on (see Fig. 1a-v and Methods). One quarter oscillation cycle later, when the atoms are centered in X (see Fig. 1a-vi), we recool them and take another picture. This third set of pictures facilitates the occupancy determination in $X_SYZ$, making the occupancy error in the state measurement lower than $10^{-4}$ (see Methods). Between the first and second images, we detect that 2.2% of the atoms are lost, which is consistent with the measured loss rate from collisions with the background gas over this time. We also see no evidence of additional loss or site hopping due to the process of transferring the atoms back to X.

With essentially lossless detection, we can re-initialize our qubits after state measurement, a procedure we demonstrate in Figure 3. We first sort to get a perfectly filled 5 × 5 × 2 pattern (see Fig. 3a)[7]. We then prepare an equal superposition of the two stretched states, execute state detection in $X_SYZ$ (see Fig. 3b), and reload into XYZ (see Fig. 3c). Atom loss to a background gas collision can occur in either stage; in this implementation, one atom was lost between state detection and reloading. We re-sort, filling the vacancy with an atom from the reservoir region (within 5 × 5 × 5 but outside 5 × 5 × 2) (see Fig. 3d). The ability to fix qubit loss errors will ultimately be an important element of quantum error correction.

Because of their insensitivity to noise, it is preferable to quantum compute in the clock states, $|F = 4, m_F = 0\rangle$ and $|F = 3, m_F = -0\rangle$[19]. However, these states do not move in the state-dependent lattice. To generalize our detection method, we use two AFP pulses to transfer atoms from superpositions of clock states to superpositions of $|F = 4, m_F = -2\rangle$ and $|F = 3, m_F = -2\rangle$, and state detect from there (see Methods). We generate a range of clock state superpositions by transferring all the atoms to the $|F = 3, m_F = 0\rangle$ and executing a $\pi/2$-$\pi$-$\pi$-$\pi/2$ sequence on the clock transition, scanning the phase of the final pulse. We do this in a one quarter depth XYZ lattice, which is preferred for minimizing decoherence while maintaining gate fidelity. Before state detection, we adiabatically raise the lattice to full power. The result is

shown in Fig. 4a, with 480 μs between the π/2 pulses. The fraction of atoms found in the wrong state at phase π is 0 $(^{+5}_{-0})$ × 10$^{-4}$, which is to say we found no errors in 2200 atom detections, consistent with as good a fidelity as we measured starting from the stretched states. The error at phase 0 is larger, 22 $(^{+12}_{-8})$ × 10$^{-4}$. We attribute the worse performance when measuring from the $|F = 4, m_F = 0\rangle$ state to Rabi frequency differences among the various AFP microwave transitions (see Methods). Adding the ability to dynamically change the microwave polarization should avoid this issue completely.

To further make use of our state detection method, we measure our qubit coherence time by adding additional intermediate π-pulses at 20 ms intervals (see Methods) and performing the spin-echo measurement described above with longer evolution times. From the observed fringe contrast as a function of time, shown in Fig. 4b, we find that the coherence times ($T'_2$ and $T_1$) for atoms in a 5×5×5 volume is 12.6 s. The improvement over our previously reported coherence time of 7.4 s[18,7] results from better cooling and a farther detuned optical lattice, both of which minimize spontaneous emission from the lattice.

The concept of our state measurement method, where internal states are mapped onto atom positions, can be generalized to atoms in reconfigurable dipole trap arrays[20-23]. Each dipole trap can be formed by overlapping two traps with opposite circular polarizations. For state detection, the two traps can be spatially separated. If needed, dipole heating could be minimized by making the two traps linearly polarized before the atoms are imaged.

While it can now be used for final state readout, we ultimately plan to adapt this measurement to quantum error correction, where state detection of only a subset of the atoms is required[24-26]. The atoms to be detected would be selectively transferred to larger $m_F$ states. During the motion, the clock state qubits will remain weakly trapped at their original position. Transferring them to $X_S$ will probably require matching the nodes of $X$ to nodes of $X_S$. Speed concerns will require replacing fluorescent detection with phase contrast imaging[27], which would act as a form of holography and allow all planes to be imaged at once. Preventing the undetected qubits from interacting with detection light will require shifting qubits out of resonance, transferring them to dark states or using a second atomic species for measurement[28]. Our state detection error approaches the commonly accepted threshold of $10^{-4}$ for fault-tolerant quantum computation[29], and it already comfortably surpasses the thresholds of $\sim 10^{-3}$ for some surface codes[30].

**Acknowledgements** This work was supported by the U.S. National Science Foundation grants PHY-1520976 and PHY-1820849.

**Author Contributions** All authors contributed to the design, execution and analysis of the experiment and the writing of the manuscript. A.K., T.-Y.W. and F.G. collected all the data.

**Competing Financial Interests** The authors declare no competing financial interests.

**Materials and Correspondence** The data presented in this report are available on request to D.S.W.

**Fig. 1**

a

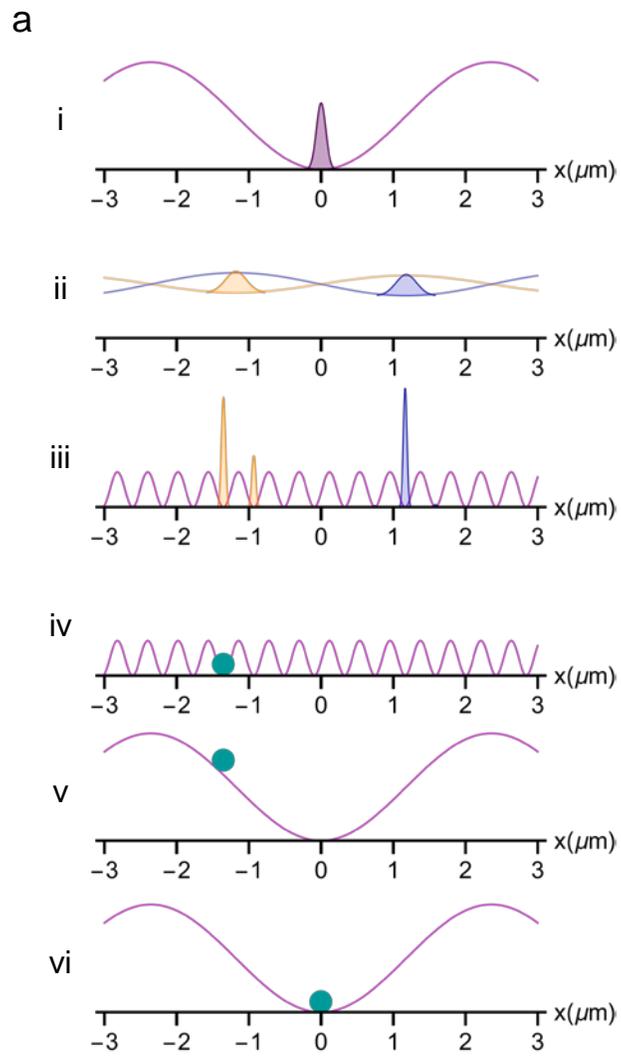

b

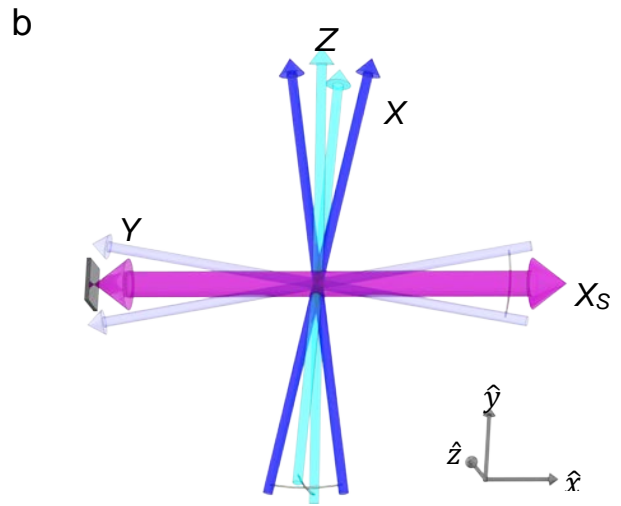

**Fig. 1 | Overview of lossless state detection. a,** State detection steps. The potential energy curves have a consistent vertical scale and the atomic wavefunctions are normalized. **a-i**: Pre-measurement state. The purple curve denotes the state-independent potential energy in the *x* direction (*X* lattice). The purple shaded region represents the wavefunction of an atom in the vibrational ground state and an equal superposition of two states. **a-ii**: Displaced state-dependent lattices. The potential has been adiabatically transformed into two shallower state-dependent potentials, where the potential energy and wavefunctions for each of the two states are shown in orange and blue respectively. **a-iii**: Transfer atoms to $X_S$. For each internal state, the number of $X_S$ lattice sites with significant wavefunction amplitude depends on the relative position of the $X_S$ and the state-dependent $X$ potential energy minima, which is not fixed in our experiment. **a-iv**: Image atoms in $X_S$. The wavefunction has been projected onto a single lattice site. The atom, now spread among many internal states and several vibrational levels, is denoted by the green ball. Its location is used for state assignment. **a-v**: Atom transferred back to $X$. **a-vi**: Atom at the bottom of $X$ after one quarter oscillation period. At this point another image is taken**. b,** Diagram of the lattice beams. The *X* (dark blue), *Y* (grey) and *Z* (light blue) lattices consist of pairs of linear-polarized laser beams crossing at 10º angles (drawn as 20º in the figure, but otherwise to scale), giving 4.8 μm lattice spacings. The beams are focused at the atoms with a Gaussian waist of 75 μm. The $X_S$ lattice (purple with double arrows) is formed by retro-reflection, resulting in a 0.42 μm lattice spacing. The $X_S$ beam has a ~150 μm Gaussian waist at the atoms.

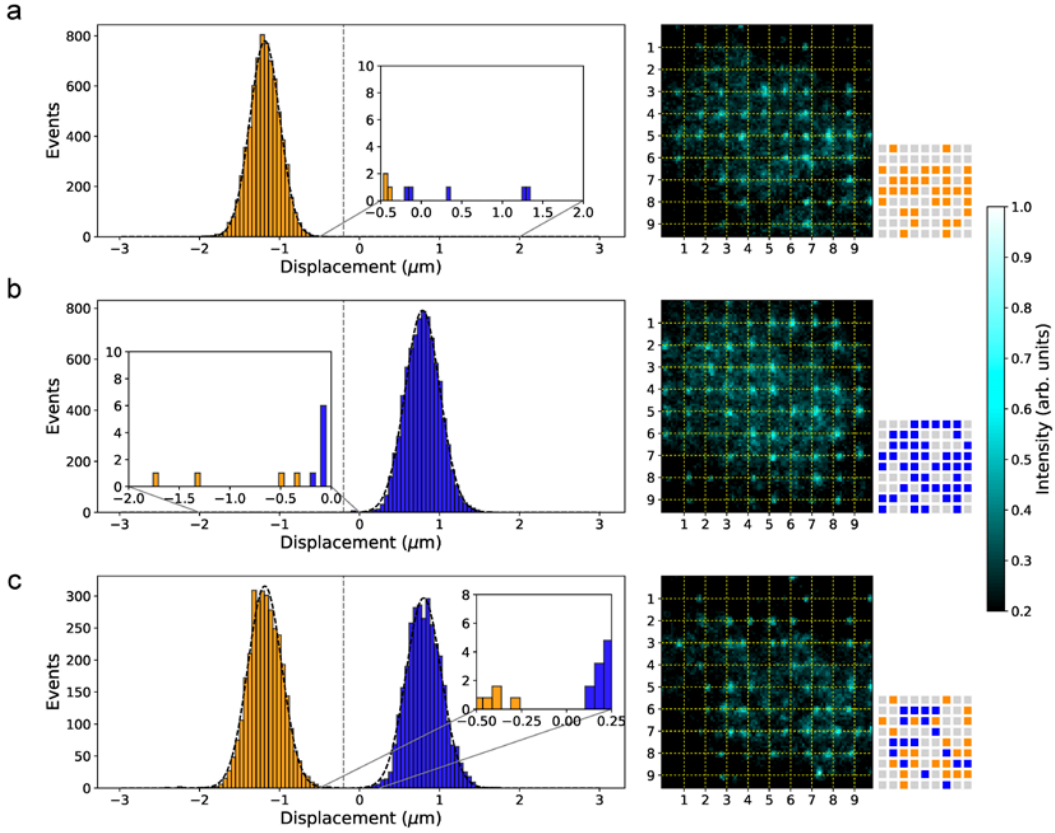

**Fig. 2 | Displacement distributions and state assignment.** The histograms show the displacement distributions during state detection. **a**, Atoms prepared in $|F = 4, m_F = -4\rangle$. **b**, Atoms prepared in $|F = 3, m_F = -3\rangle$. **c**, Atoms prepared in a superposition state. For each histogram we perform 50 implementations starting with a 30-40% randomly loaded $9 \times 9 \times 5$ lattice. The black dashed lines are Gaussian fits to the distributions. The vertical grey dashed line at -0.2 μm sets the dividing line for state assignment. Atoms to the left (right) are ascribed to $|F = 4, m_F = -4\rangle$ ($|F = 3, m_F = -3\rangle$) and colored orange (blue). The state detection errors, those events that are nominally prepared in one state but are detected in the other state, are highlighted in the insets of **a** and **b**. An example image of one lattice plane is shown next to each histogram. The color bar to the right is the intensity key, which uses contrast enhancement. The displacement of each atom in the *x* direction (horizontal in

the image) is the difference between its fit center and its initial position, which is marked by the dashed yellow grid. The occupancy map for each example image is shown in the associated square pattern. Orange or blue denotes an occupied site in one of the two states, while grey represents an empty site.

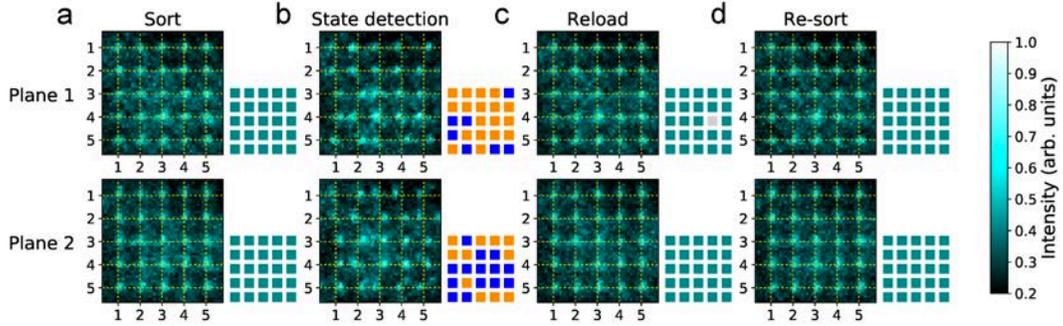

**Fig. 3 | Demonstration of re-initialization of a 3D qubit array.** The two images in each column correspond to different lattice planes. The color bar to the right is the intensity key, which has contrast enhancement. The dashed yellow grid marks the initial atom locations. The associated square patterns show the occupancy maps. Cyan denotes an occupied *XYZ* site, orange and blue denote an atom determined to be in either the $|F = 4, m_F = -4\rangle$ or $|F = 3, m_F = 3\rangle$ state respectively, while grey denotes an empty site. **a**, Perfect filling of a $5 \times 5 \times 2$ array after sorting from a randomly half-filled $5 \times 5 \times 5$ *XYZ* lattice. **b,** Lossless state detection in $X_S YZ$ after the atoms were prepared in an equal superposition of $|F = 4, m_F = -4\rangle$ and $|F = 3, m_F = -3\rangle$. **c,** Atoms reloaded into *XYZ*. Note that an atom in Plane 1 (at (4,3)) has been lost due to a background gas collision. **d,** Resorted atoms. The vacancy has been filled with an atom from the reservoir region (outside these two planes but within $5 \times 5 \times 5$), where extra atoms had been left during the initial sorting.

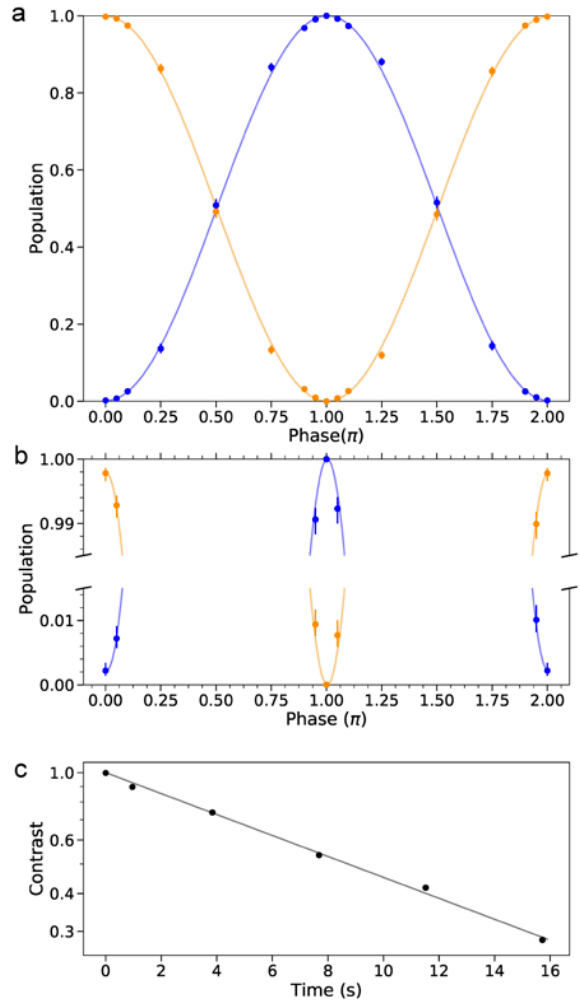

**Fig. 4 | State-selective detection from the clock states. a**, Fringes from a spin-echo measurement sequence on the $|F = 4, m_F = 0\rangle$ and $|F = 3, m_F = 0\rangle$ transition. The evolution time is 480 µs and the phase of the final π/2 pulse is scanned. Atoms are transferred with microwave AFP pulses to $|F = 4, m_F = -2\rangle$ (orange) and $|F = 3, m_F = -2\rangle$ (blue) for state detection. At each point, the populations of both states are determined in the same state measurement. The data points at phase 2π are the same as those at phase 0. The solid curves are fits to a sine function, and the fit amplitude gives the fringe contrast. **b**, Magnified regions of **a**. At phases π and 0, perfect state preparation and detection would lead to all atoms being in the same state. At π-phase, there were no errors among the 2200 atoms we measured. We attribute the $2\times10^{-3}$ error at 0-phase to the lower quality of some AFP pulses (see Methods). Error bars in **a** and **b** represent one standard deviation and are due to

counting statistics. The error bars for the data points at phase $\pi$ are smaller than the size of the symbols. **c**, Semi-log plot of fringe contrasts for spin-echo sequences, for a range of evolution times. The fit exponential time constant, which gives the coherence time of the qubits, is 12.6 s. The fit error bars are smaller than the size of the symbols.

## Methods

### Occupancy map generation in the *XYZ* lattice

The location of the *XYZ* lattice sites with respect to the camera is actively locked[18]. Around a site center we integrate a 3-pixel-wide region in the *y* direction and use 11 pixels (one pixel = 0.52 μm in real space) in the *x* direction to perform a least-squares fit to a Gaussian with an offset, keeping the center and the width fixed. When the fit amplitude is larger than a predetermined threshold, we preliminarily mark the site as occupied.

Using the preliminary occupancy map, we sort the atoms into four occupancy cases: no adjacent atoms in the z-1 or z+1 plane, an adjacent atom in the z-1 plane, an adjacent atom in the z+1 plane, and adjacent atoms in both the z-1 and z+1 plane[9]. We use a random half lattice filling in order to build statistics for each occupancy case. Each case has a different threshold, determined from clear gaps in the distribution of fit amplitudes from thousands of site measurements. For each edge plane, we can modify the thresholds to account for the uncertainty caused by the unmeasured adjacent plane, but the occupancy error is still roughly doubled for those planes. The occupancy assignment for all sites over the five planes is iterated at most five times, to ensure that they go to the right occupancy cases. For a single set of images in *XYZ*, we achieve histograms like those in Supplementary Fig. 1a, which allow us to discriminate the occupancy for a site in the center three planes with an error of $3 \times 10^{-4}$. When we combine the information from two sets of images of the same atoms, as in the sets taken before and after the state detection, we can decrease the error to $8 \times 10^{-5}$.

### Occupancy generation in the $X_S YZ$ lattice

When analyzing images in the $X_S YZ$ lattice, we use the *XYZ* centers as reference points for the displacements. Around initially occupied sites, we integrate a 3-pixel-wide region in the *y* direction, search for the highest intensity among 8 pixels in the *x* direction, and select 7 out of these 8 pixels to perform a least-squares fit to a Gaussian with variable amplitude, offset and center and a fixed width. If the site is still occupied, we use the fit center to determine its state. Using fewer pixels makes the fits less likely to chase background noise or be disrupted by the intensity tails of neighboring atoms. We apply the fixed-center Gaussian fit to initially unoccupied sites.

The fit amplitude histogram for the occupancy case of no atoms in the adjacent $z$-planes is shown in Supplementary Fig. 1b. The left (right) peak corresponds to the unoccupied (occupied) sites. The lack of a clear gap between the two peaks is because floating center fits have a higher variance. We address this problem by taking a final set of images in *XYZ*. The blue peak in Supplementary Fig. 1c shows the state detection amplitudes for sites at which the final image had an atom. Sometimes an atom is there during state detection, but is lost before the final image. To distinguish this possibility from the atom being lost between the first image and the state detection, it is useful to plot the amplitudes for initially empty sites fit with a floating center at state detection (the red peak in Supplementary Fig. 1c), since this is the amplitude distribution at sites that have lost atoms. The overlap between these two histogram peaks makes it impossible to always distinguish between these two possibilities, but as we will now explain, we can substantially minimize the impact of this effect on our measurement.

Our logic is summarized in Supplementary Fig. 2. If the atom is in the final image, it was there for state detection. If it is not in the final image, then it was lost either before the state detection or after the state detection. We assign a set of thresholds (one for each occupancy case) to the state detection histograms, placing them at the upper edge of the tail of the histogram of empty sites (the dotted line in Supplementary Fig. 1c). If the amplitude is above the threshold in the state detection image, we infer that there was an atom during state detection that was lost afterwards, and its position is counted in the state measurement. If it was below the threshold, we place it in the category of lost before state detection. Since 19% of atoms at state detection have an amplitude below the threshold, and ~1.1% of them are subsequently lost, this procedure artificially inflates our loss before the state detection by 0.2%, equivalent to a 10% worse background gas pressure. This pseudo-loss is not inherent to the detection scheme; were the loss to be reduced, the pseudo-loss would be proportionally reduced.

Only false positive identifications during state detection lead to state detection errors.

When an actually empty site is falsely identified as occupied, the floating-center Gaussian fit renders an arbitrary fit center that could be ascribed to either state. Given our choices of thresholds, we infer that this error occurs $3.6 \times 10^{-5}$ of the time. There is an additional contribution of $2.9 \times 10^{-5}$ from the far end of the red distribution in Supplementary Fig. 1C. The contribution to state detection error from occupancy misidentification is thus only $6.5 \times 10^{-5}$. This is consistent with the our state detection measurements (as in Fig. 2a and Fig. 2b), where of the 9 errors we see out of 16000 state detections, one seems to be due to a site occupation error.

**Phase scrambled optical molasses**

Our PGC imaging light consists of three pairs of retro-reflected molasses laser beams, forming a 3D standing wave pattern that is not interferometrically stable. Some of the variation in detected imaged amplitudes results from variations in the cooling light intensity at the center of each lattice site. The effect is more pronounced in $X_SYZ$ than in $XYZ$, since the cooled atoms are more localized in $X_SYZ$, so the intensity variations of the PGC light are averaged over a smaller volume.

To address this issue, we installed liquid crystal phase modulators at the input of two of the three molasses pairs. We apply linear phase ramps of $2\pi$ over 30 ms and $\pi$ over 33 ms in the two arms. The phase space of the interference pattern is thus sampled broadly over the course of the 150 ms imaging time per plane. The phases are changed slowly enough that they do not compromise the PGC. The histogram we obtain, which is the one shown in Supplementary Fig. 1b, has about a 20% narrower right side peak than similar histograms obtained without this phase modulation.

**Atom reloading from the $X_SYZ$ lattice to the $XYZ$ lattice**

After state detection, $X$ with regular polarization is turned on with a 2 ms ramp. Then $X_S$ is first ramped down from 98 kHz trapping frequency to 43 kHz in 5 ms and then suddenly shut off. The cold atoms are thus left about half way up the hill of the $X$ lattice trap. They fall to the trap center in about one quarter oscillation cycle (17 μs). We then polarization gradient cool to stop the atoms near the trap bottom, and take the final image set in $XYZ$.

**Asymmetric displacement centers for the two stretched states**

We have measured the displacement centers for both stretched states using a range of applied voltages on the Pockels cells (see the data points in Supplementary Fig. 3). If the polarization were perfect, we would expect the motion of atoms in the two states to follow the dashed lines, which have similar trajectories for the two states. But we see an asymmetry in this data, which can also be seen in the difference between the displacements of the two states in Fig. 2. The intensity of non-linearly polarized light for the $X$ beams before the vacuum cell is smaller than 0.0007. However, it is possible that the two linear polarizations are not exactly perpendicular to the propagation plane of the $X$ light, and it is possible that cell birefringence compromises the polarization. To get an idea of how polarization imperfections might affect these displacements, we calculated the expected motion with a fixed 1% relative intensity of perpendicular $\pi/2$ out-of-phase linear polarization in the $X$ beam whose polarization is not rotated (the solid lines). While the imperfect match to the data makes it clear that this is not the specific polarization flaw in $X$, the improved agreement with the observations support the idea that polarization imperfections underlie this discrepancy.

**State detection in the two $m_F = -2$ states**

Since the two $m_F = 0$ qubit states do not move in the state-dependent lattice, after the spin-echo measurements we transfer them toward the stretched states in order to generalize our state detection protocol. The higher $|m_F|$, the deeper the state-dependent lattice, and thus the less spontaneous emission there is. But it takes more AFP microwave pulses to get to higher $|m_F|$, so the optimal $|m_F|$ depends on the AFP pulse fidelity. For our current microwave polarization the $|F = 3, m_F = -2\rangle$ to $|F = 4, m_F = -3\rangle$ transition has a 0.29 times lower Rabi frequency than the $|F = 4, m_F = -2\rangle$ to $|F = 3, m_F = -3\rangle$ transition. A single AFP pulse optimized for the latter transition does not drive the former transition efficiently. Therefore, for now, we stop at the $m_F = -2$ states. In order to account for the shallower potentials, we increase the voltage ramping time from the 250 μs we use for the stretched states to 500 μs.

The fact that the error is worse at phase 0 in the short time spin-echo measurement (see Fig. 4) has a similar cause, a 0.7 times lower Rabi frequency for the $|F = 3, m_F = -1\rangle$ to |F = 4, $m_F$ = -2⟩ transition compared to the |F = 4, $m_F$ = -1⟩ to |F = 3, $m_F$ = -2⟩ transition. The parameters that optimize the second AFP pulse for the latter transition are not optimal for the former. Some atoms are left behind in the |F = 3, $m_F$ = -1⟩ state, from which they are detected to be in the wrong state. Using a dual-polarized horn antenna would enable the microwave polarization to be dynamically changed, which would allow the two Rabi frequencies driven by each AFP pulse to be equalized.

**Spin-echo measurements with dynamical decoupling**

Before starting the spin-echo measurements, we lower the *XYZ* lattice to one quarter depth and use five AFP pulses to transfer the vibrationally cooled atoms from |F = 4, $m_F$ = -4⟩ to |F = 3, $m_F$ = 0⟩. To ensure that all atoms start in |F = 3, $m_F$ = 0⟩, we apply a clearing pulse to push away any residual |F = 4⟩ atoms. We then apply an AFP pulse to transfer to |F = 4⟩ atoms that might have been left behind in |F = 3, $m_F$ = -1⟩ and |F = 3, $m_F$ = -2⟩, and another to empty |F = 3, $m_F$ = -3⟩. Then we clear |F = 4⟩ again. These precautions may not be necessary, since the optimized AFP pulse fidelity might be better than our empirical lower limit of 0.999.

Our spin-echo sequences start with a 60-μs π/2 pulse that puts the atoms in equal superpositions of |F = 3, $m_F$ = 0⟩ and |F = 4, $m_F$ = 0⟩. For the spin-echo measurement of Fig. 4a, with a 480 μs total evolution time, the times between successive pulses in the π/2-π-π-π/2 sequence are 120 μs, 240 μs and 120 μs. For longer measurement times, we apply a 120-μs long dynamical decoupling π-pulses every 20 ms after the first 10 ms. We chose 20 ms to balance the effects of inhomogeneous broadening, lattice intensity drifts and magnetic field drifts (which all favor a shorter time between the pulses) and the effect of drift in the microwave pulse power (which favors fewer pulses). We use an even number of π–pulses, alternately phase-shifting them by 180°, starting from in phase with the initial π/2 pulse. After a variable amount of time we end the sequence 10 ms after the last π–pulse with a final π/2 pulse, scanning its phase from 0 to 2π.

**The effect of spontaneous emission from the lattice on state detection**

Decoherence in this system is mostly due to the spontaneous emission of lattice light,

which scrambles the phase of atoms in the qubit states, but also transfers atoms out of the qubit basis into other magnetic sublevels. Most of these will show up in one of the state detection peaks. Those that end up in the $|F = 3, m_F = -1\rangle$ and $|F = 4, m_F = -1\rangle$ levels are special cases, since the first of the final two AFP pulses will transfer them back to the $|F = 4, m_F = 0\rangle$ and $|F = 3, m_F = 0\rangle$ states, which do not move in the otherwise state-dependent lattice. We have checked this by performing the state detection method without the AFP pulses; we measure a single peak centered near 0 μm. As lattice scattering starts to accrue, the empty region between the two peaks in the displacement distributions (Fig. 2c) starts to fill. These atoms, like all atoms that undergo spontaneous emission, degrade the contrast. Since they preferentially lie on the $|F = 3, m_F = 0\rangle$ side of the state dividing line at -0.2 μm, they lead to a uniform upward shift of the fringe, which is inconsequential from the point of view of the coherence measurement.

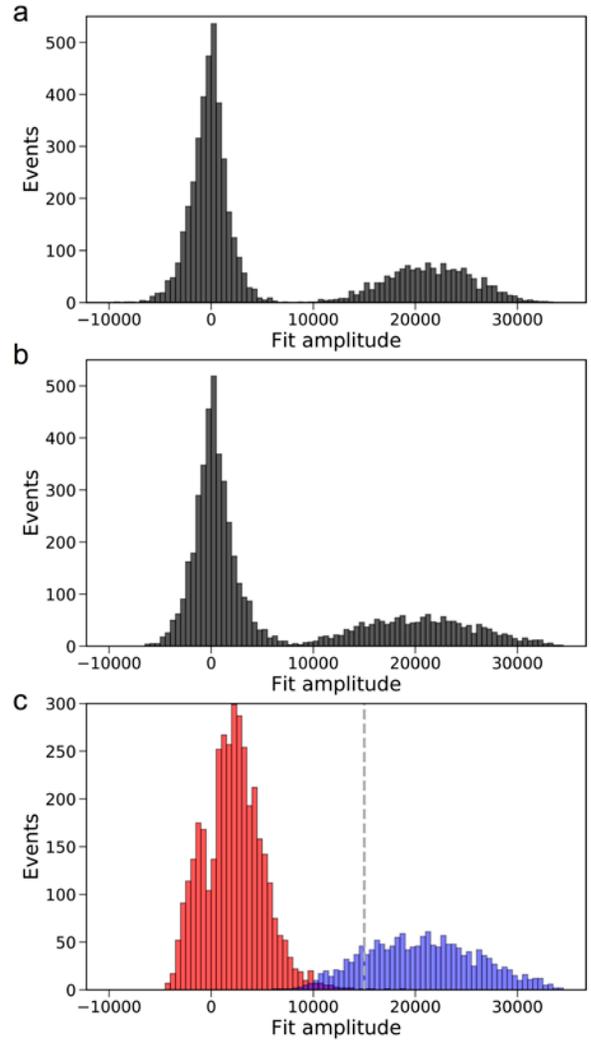

**Supplementary Fig. 1 | Fit amplitude histograms.** The histograms are derived from the same data that yielded Fig 2a. They are for the occupancy case with no atoms in the adjacent z-planes. The other occupancy cases have similar sets of histograms. **a**, Fixed center fit on every site for the initial picture in the *XYZ* lattice. The empty and occupied peaks are well separated. **b**, Data from state detection in the $X_SYZ$ lattice. Initially empty sites are fit with fixed center peaks and initially occupied sites are fit with floating centers. There is less of a gap between the peaks than in **a**. **c**, Occupancy analysis in $X_SYZ$ using additional information. The red points come from floating center fits of initially empty sites, showing how the fitting works at sites that lose an atom before the state detection. The distribution is broader than fixed center fits of empty sites, with more of a high amplitude tail. The blue points are due to only atoms that are still there in the final *XYZ* images. The

vertical dashed line is the threshold we set to be above almost all red points. If there is no atom in the final image, then we assume that any state detection amplitude below the threshold corresponds to an atom that was lost during the state detection.

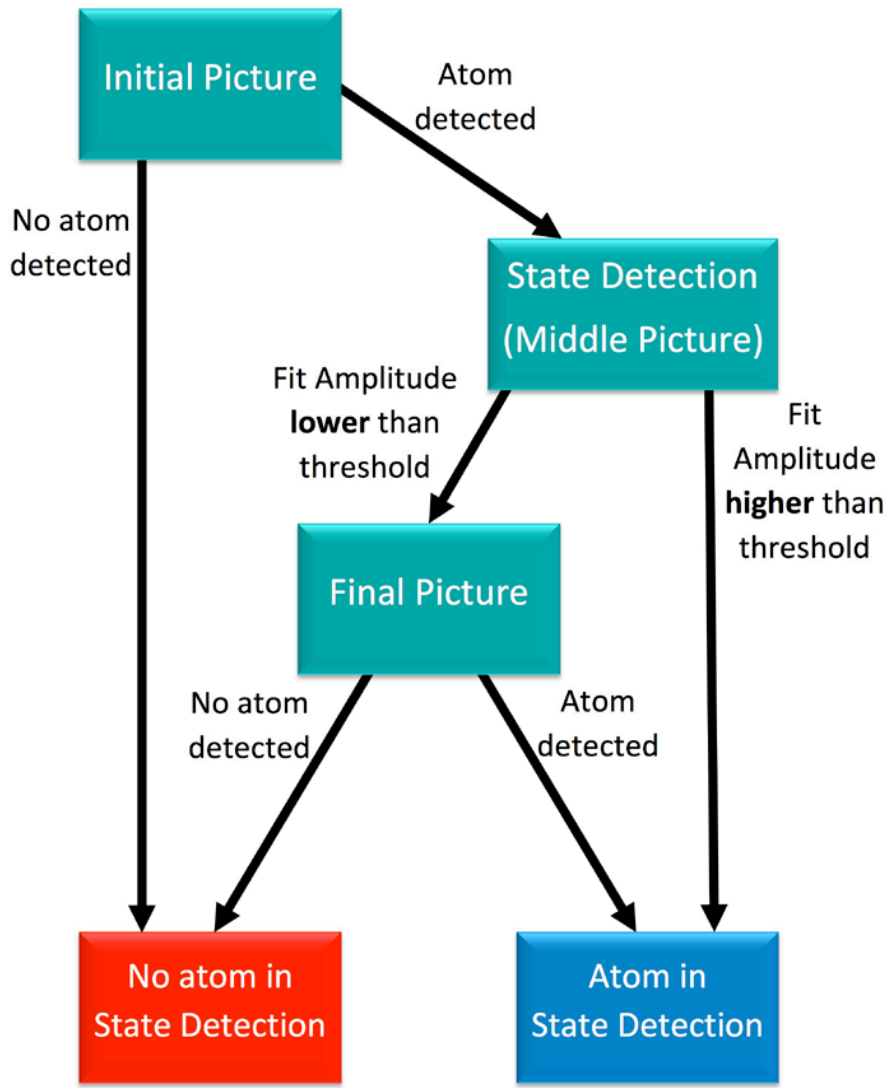

**Supplementary Fig. 2 | Occupancy assignment flow chart for state detection.** There is a similar flow chart with different thresholds for each occupancy case. The end result is a contribution of $<7\times10^{-5}$ to the state detection error, and a ~10% virtual increase in the number of atoms lost to background gas collisions, which amounts to 0.2% of the atoms given the quality of our vacuum.

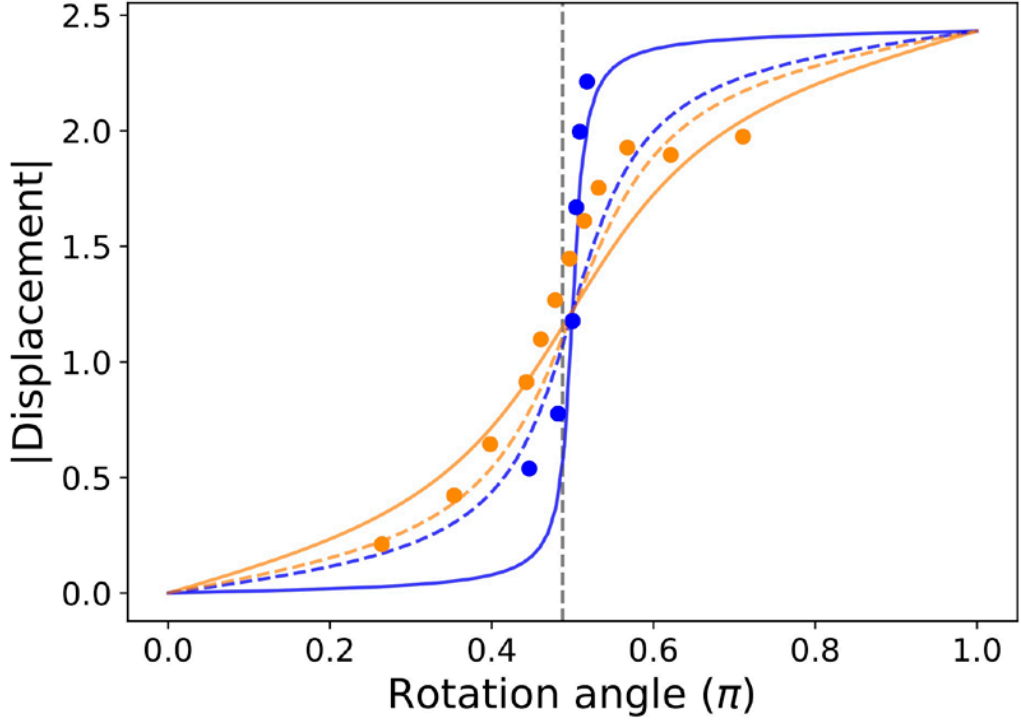

**Supplementary Fig. 3 | Displacement magnitudes as a function of polarization rotation angle.** The orange (blue) dots are extracted from the state detection measurements with atoms prepared in $|F = 4, m_F = -4\rangle$ ($|F = 3, m_F = -3\rangle$) for various final polarization rotation angles of one of the two *X* lattice beams. The fit error bars are all smaller than the symbol sizes. The dashed curves are from calculations using ideal polarizations. The solid curves assume a fixed 1% relative intensity of perpendicular $\pi/2$ out-of-phase linear polarization in one *X* beam. These curves support our assumption that the observed asymmetry in displacement we see is due to some sort of modest polarization imperfection. The vertical dashed line is at the angle at which we take the data for Figure 2.